\documentclass[12pt]{iopart}

\usepackage{graphicx}
\usepackage{amssymb}
\usepackage{graphics}
\usepackage{epsfig}

\begin{document}

\title{Entropic contributions to the splicing process}

\author{Matteo Osella, Michele Caselle}
\address{Dipartimento di Fisica Teorica and INFN, Universit\`a degli Studi di Torino, v. Pietro Giuria 1, 10125, Torino, Italy
  \ead{mosella@to.infn.it ~caselle@to.infn.it}}

\date{\today}

\begin{abstract}
 It has been recently argued  that the depletion attraction may play an important role in different aspects of the cellular organization, 
 ranging from the organization of transcriptional activity in transcription factories to the formation of the nuclear bodies. In this paper 
 we suggest a new application of these ideas in the context of the splicing process, a crucial step of messanger RNA maturation in Eukaryotes. 
 We shall show that entropy effects and the resulting depletion attraction may explain the relevance of the aspecific intron length variable in 
 the choice of the splice-site recognition modality. On top of that, some qualitative features of the genome architecture of higher Eukaryotes can find an
  evolutionary realistic motivation in the light of our model.

\end{abstract}


\bigskip


\maketitle

\section{Introduction}
\label{intro}
\subsection{The splicing process and the spliceosome assembly}

Eukaryotic genes have a split nature, in which the exons, that encode the information for the final product 
of a messanger RNA (mRNA), are interrupted by non-coding introns in the DNA and in the precursor mRNA (pre-mRNA) transcript.  
The intron excision and the concomitant joining of exons, which basically represent the splicing process, are a necessity in order 
to obtain a mature mRNA that can be exported in the cytoplasm and for example correctly translated into a protein.  
This process is  carried out by the spliceosome, a macromolecular ribonucleoprotein complex,  that  assembles on pre-mRNA  
in a stepwise manner \cite{macmillan,wahl,black}. The first requirement is the correct recognition of the intron/exon boundaries by small nuclear 
ribonucleoproteins (snRNPs) 
and some auxiliary splicing factors by binding to sequences located at the ends of introns. Subsequently the splice-site pairing takes place,  bringing  the two
 exons near to each other and  looping  the intron that have to be cut away.  
\subsection{Exon definition and intron definition}

\begin{figure}[h]
\begin{indented}
\item[]\includegraphics[width=8cm]{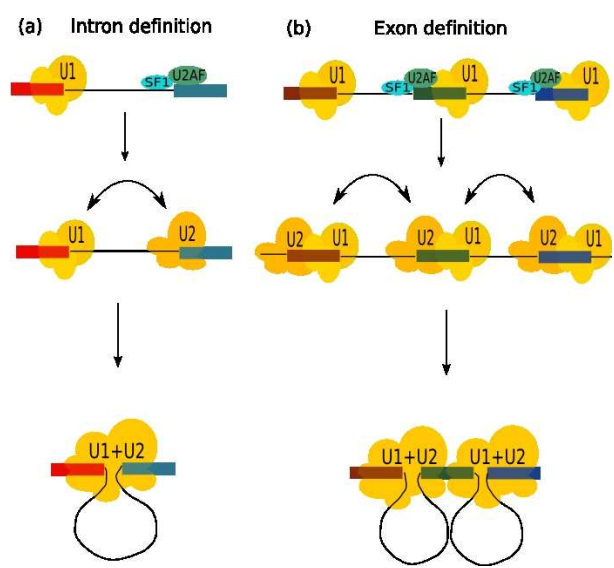}
\end{indented}
\caption{Intron definition and exon definition: two ways of splice-site recognition.}
\label{fig1}
\end{figure}

Although the molecular players and the key steps of spliceosome assembly are remarkably conserved through different species \cite{collins}, 
there are two alternative pathways of splice-site recognition: \textit{intron definition} and \textit{exon definition} \cite{macmillan,berget,ast,ram,hertel}.\\
 Intron definition (see figure \ref{fig1}a) begins with the direct interaction of the U1 snRNP with the splice-site in the upstream end of the intron 
 (5' splice-site). 
The splice-site in the downstream end (3' splice-site) is then recognized by the U2 snRNP and associated auxiliary factors such as U2AF and SF1.  
When the two complexes are constructed on the intron/exon boundaries they can be juxtaposed, closing an intron loop 
which is then spliced away in order to correctly glue the exons. The interaction of the splicing factors bound at 
the splice-sites occurs in this case across the intron. The exon definition (see figure \ref{fig1}b) requires 
instead that the initial interaction between the factors bound at the splice-sites occurs across the exon: 
the U1 and U2 snRNP and associated splicing factors bind to the 3' and 5' ends of an exon and a complex is built 
across it (usually with the participation of serine/arginine-rich (SR) proteins \cite{ast}); then complexes on different exons join together 
so as to allow intron removal.

\begin{figure}[h]
\begin{indented}
\item[]\includegraphics[width=0.8\columnwidth]{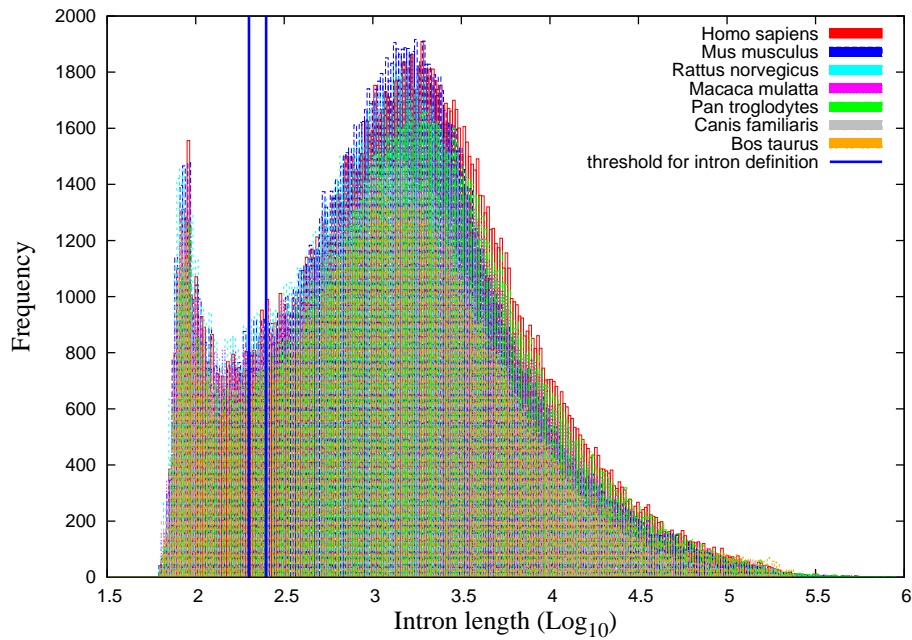}
\end{indented}
\caption{Intron length distribution for different higher Eukaryotes. The  distribution shows a two peaks
structure which is remarkably universal. The intron length threshold mentioned in the main text (blue lines) is located exactly 
between the two peaks. The right peak contains mostly introns which undergo exon-defined splicing, while the left one can be associated to 
intron-defined spliced introns. The coordinates of introns used were downloaded from the Ensembl database vers.47 \cite{ensembl} 
and the distribution was obtained through a logarithmic binning.}
\label{intron-distribution}
\end{figure}

Previuos studies have shown that the length of the intron that has to be removed has a key role in the choice 
of the splice-site recognition modality \cite{ram,hertel2,collins2}.
Short introns are spliced away preferentially through intron definition, while longer introns seem to 
require an exon definition process. In particular the analysis of \cite{hertel2} suggests
 the presence of a threshold in intron length - between 200 and 250 nucleotides (nt) long-  above which 
intron-defined splicing ceases almost completely. Lower Eukaryotes present typically short 
introns, below the threshold, so it is expected that intron removal proceeds through intron 
definition \cite{berget,hertel,collins2,mcguire}. Higher  Eukaryotes instead have an  intron length distribution presenting two 
pronounced peaks, with the threshold in between (see figure\ref{intron-distribution}), so even 
if the vast majority of introns are above the threshold (data in table \ref{tabella}), the 
first peak contains introns suitable for intron definition.
This agrees with several studies \cite{collins2,mcguire,lim,sakabe} which have shown that both ways of splice-site 
recognition are present in higher Eukaryotes, even if the exon definition pathway seems to be the prevalent one.

As it can be seen in figure \ref{intron-distribution}, not only the shape of the distribution is quite conserved through different species, 
but also the position of the peaks.

\section{Intron removal and depletion attraction}
\label{sec:1}
The first goal of our paper is to propose a simple physical model of early steps of spliceosome assembly on a pre-mRNA, taking 
into account possible entropic contributions to the splicing process. Subsequently we will show that, despite its simplicity, the 
model is able to produce quantitative predictions which are in rather good agreement with experimental and bioinformatical observations.\\
Our starting point is the assumption that the splicing complexes, which are immersed in the crowded nuclear
environment (\cite{marenduzzo2} and reference therein), feel the so called ``depletion attraction''~\cite{AO}. This interaction is 
essentially an entropic effect due to the fact
that when two large complexes (like the splicing ones) approach each other, they reduce the volume between them excluded to the depleting particles.
 If the complexes are immersed in an environment crowded of macromolecules of
smaller (but comparable) size, then this excluded volume effect
induces an attractive interaction between the two complexes. 

This simple geometrical reasoning forms the basis for the Asakura-Oosawa (AO) theory \cite{AO,AO2}. In more recent years, a  more sophisticated
hypernetted-chain-based theory describing depletion forces in fluids  has been developed \cite{attard1,gotzelmann} and tested in Monte Carlo simulations
\cite{attard3}. However, as discussed in \cite{marenduzzo2}, the AO theory is an approximation that remains quite accurate up to $c \sim 0.3$, 
with $c$ representing
the fraction of volume occupied by the crowding molecules. As far as the $c$ value  inside a living cell has been estimated between 0.2-0.3 \cite{ellis,minton} 
we can safely use in the following the simpler AO description of depletion effects.

Since the two splicing complexes are joined by a freely
fluctuating RNA chain the depletion-based interaction becomes effectively  long range, with a logarithmic 
dependence on the chain length. We suggest that this depletion attraction is the driving force which allows the splicing complexes to meet and
join one another, in order to start up the splicing process. As we shall see
this assumption naturally leads to a smooth cross-over from an intron defined to an exon defined splicing pathway
as the chain length increases.

\subsection{Presentation of the model}
\label{presentation}

Let us model, as a first approximation,  the pre-mRNA as a Freely Jointed Chain (FJC) \cite{cantor,flory}, i.e a succession of
  infinitely penetrable segments, each of length $l$ equal to the Kuhn length of the single strand RNA (ssRNA). 
The estimated Kuhn length of ssRNA is approximately in the range 2-4 nm, i.e 3-6 nt \cite{rippe,liphardt,mills}. We chose to neglect the 
self avoidance in order to use the analytical tractable FJC and moreover the diameter of ssRNA, approximately 2 nm, is not so relevant with 
respect to long chains:  as reported in \cite{rippe,rippe2} the FJC modelization is suitable for ssRNA chains with a length greater than 5-6 Kuhn 
segment, as will always be the case in the following.  

 The two complexes, composed by U1, U2 and
  splicing factors, that bind to the exon/intron
  boundary in the intron definition process, will be  modeled as spheres with a diameter $D$ (the
  dimensions of the major components U1 and U2 are quite similar, both of the order of $\sim 10$~nm,
  see \cite{U1} and \cite{U2} for details).
   The same geometrical approximation will be done for complexes constructed across exons in exon definition.
   They will be considered as spheres of diameter $D'$, with  $D' \sim 2D$ since
   they are composed by both the U1 and the U2 subcomplexes plus the exon in between, usually with SR proteins bound to it \cite{hertel}.\\
The simple FJC model allows the analytical calculation of the radial probability distribution of the end-to-end distance \cite{cantor}:

\begin{equation}
W(r)dr = \left(\frac {\beta} {\sqrt{\pi}}\right)^{3} 4 \pi r^2 \exp(-\beta^2 r^2)dr
\label{radial-distribution}
\end{equation}

where $\beta=\left(\frac 3{2nl^2}\right)^\frac1{2}$ , $n$ is the number of indipendent segments in the FJC and $l$ is the length of a 
segment (in our case the Kuhn length of mRNA).
Following \cite{marenduzzo}, in order to include the depletion attraction contribution, we weighed the radial
  probability distribution of the end-to-end distance (we assume that the ends of the intron can be considered as
  the center of the beads) with a Boltzmann factor, which takes into account the
  depletion attraction potential and which is non-zero in the range $~D\leq r\leq D+d$. 
This potential is easy to  evaluate in this ``hard sphere'' approximation
  (see for instance \cite{AO2}) and takes a particularly simple expression in the $d<<D$ limit.
 We can therefore define a new function $W'(r)$ as the weighed FJC radial probability distribution:

\begin{equation}
W'(r)dr = W(r)\exp(\frac3{2}c\frac D{d}(\frac {D+d-r}{d})^2)dr
\label{Boltzman-factor}
\end{equation}

where $c$ denotes the volumetric concentration of the small molecules and $d$ their typical size.
With the typical values of these quantities for the nuclear environment: $c\sim 0.2$ and $d\sim 5 nm$ one finds
for the problem at hand a potential energy of the order of one hydrogen bond, which is exactly in the range of energies needed to join 
together the two ends of an intron of length of about 10 Kuhn length (equivalent to 50 nucleotides).

Passing to the variabile $x=r-D$ (distance from the surfaces of the two spheres),  we construct our probability distribution $f(x)$ as:

\begin{equation}
f(x) =\left\{
	\begin{array}{ll}
		0 & \textrm{if }  x<0\\
		\\
                W(x)\exp(\frac3{2}c\frac {D}{d}(\frac {d-x}{d})^2) & \textrm{if } 0\le x < d \\
		\\	
		W(x) & \textrm{if }  x\ge d
	\end{array} \right.
\label{probability-distribution}
\end{equation}

which can be simply normalized as:

\begin{equation}
g(x)= \frac {f(x)}{\int_{0}^{\infty} f(x)dx}
\label{probability-normalized}
\end{equation}

It's now straigthforward to define the looping probability as the probability of finding the surfaces of the two  beads 
at the end of the chain within a sufficiently short distance $a$ (choosen as 5 nm in the following, in line with \cite{marenduzzo}):

\begin{equation}
P(x < a)= \int_{0}^{a} {g(x')dx'}
\label{loop-prob}
\end{equation}

We reported the equations for the case  $D>>d$ for the sake of simplicity, but in the numerical estimates reported in the following 
sections we used the full effective potential of depletion attraction taken from \cite{AO}.

The appealing feature of this model is that it introduces in a natural way a logarithmic relation between the intron length and the 
dimensions of spliceosome
  subcomplexes attached to its ends, if we constrain the system to keep a fixed looping probability. 
\begin{figure}[h!]
\begin{indented}
\item[]\includegraphics[width=0.8\columnwidth]{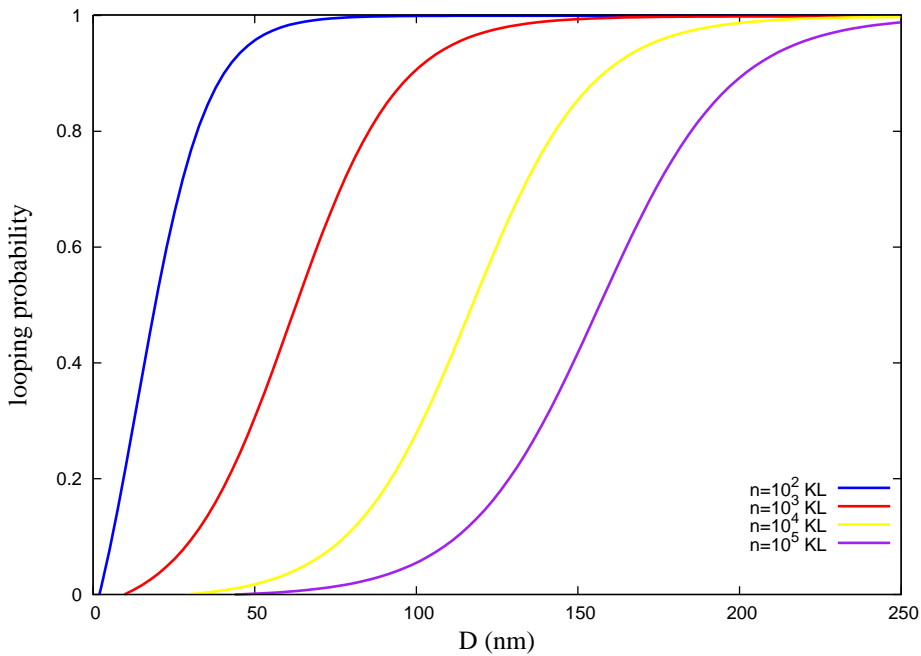}
\end{indented}
\caption{We show the looping probability (equation \ref{loop-prob} with $ a = 5nm$)  for different intron lengths as a function of the diameter of the
spheres attached to the ends. Following~\cite{rippe,liphardt} the Kuhn length of the chain was fixed to 5 nt (about 3nm). However it
is well known
that many regulatory proteins can be bound to the pre-mRNAs and that the latter may fold into rather complex secondary structures. Both these factors have 
the effect of increasing the
stiffness of the pre-mRNA thus increasing its Kuhn length. Unfortunately
so far there are no experimental estimates of the Kuhn length in these conditions, so the value derived for ssRNA should be better considered as a lower 
bound. The diameter of the small crowding molecules is assumed as 5 nm (see \cite{marenduzzo} and references therein).}
\label{loop-prob-distribution}
\end{figure}
This can be seen  by looking at figure \ref{loop-prob-distribution} where we plotted
  the looping probability for different intron lengths as a function of the diameter of the spheres attached to its ends. 
  If we increase the intron length of an order of magnitude
  the beads' diameter must be enlarged by a (roughly constant) multiplicative factor  in order
  to obtain the same looping probability.
This observation may be used to explain the switch from intron to exon definition as the intron length
increases.
  When the intron length becomes too large the dimensions of merely U1 and
   U2 subcomplexes is not sufficient to ensure a reasonable looping probability. This does not mean that such a
   process is forbidden but simply that it would require much longer times. For large enough introns it
   becomes  more probable that
   the two complexes instead join across the exon (a process mediated again by the depletion attraction), if it is sufficiently short.
   The complexes constructed across exons can actually result large enough to maintain a suitable looping
   probability, even in the case of long introns.

\subsection{Towards a more quantitative model: a compromise between soft and hard hypothesis}

Looking at figure \ref{loop-prob-distribution} we see that while the model works nicely from a
qualitative point of view it predicts intron lengths which are slightly smaller than those actually observed. In fact, in order to make the model more realistic
  and to be able to obtain also a quantitative agreement
  with the data, we must take into account two other ingredients. The first one is that
pre-mRNAs can be bound to various regulatory proteins which have the effect of increasing their Kuhn
  length. Unfortunately no direct
  estimate of the Kuhn length in this conditions exists, thus to obtain the  curves reported in
  figure \ref{loop-prob-distribution} we were compelled to use the Kuhn length of pure ssRNAs. Hence the intron length reported in the figure should be 
  better considered as lower bounds.

The second one is that the splicing (sub)complexes are rather far from the hard sphere approximation.  If the irregular shape of the molecules allows a 
snugly fit or if parts of the two subcomplexes can intermingle, the free energy gain will be larger. Again this suggests that our results should be better 
considered as lower
bounds. In this case however we can slightly improve our model and obtain also a reliable upper bound for
our looping probability.
  The maximal relaxation of the hard hypothesis can be achieved considering that the two spheres can fuse with volume conservation (\textit{soft hypothesis}).
While we can't actually write the analytical expression of the potential in this ``soft beads'' case, it's undemanding to calculate the free energy gain obtained 
by the complete fusion of the two spheres. It's directly related to the portion of volume that becomes available to the crowding molecules:

\begin{equation}
\Delta F_{gain} = cK_{B}T\left(\frac{2(D+d)^{3}-(2^{1/3}D+d)^{3}}{d^{3}}\right)
\label{soft gain}
\end{equation}
Following again~\cite{marenduzzo} we may at this point
assume that the functional dependence on $r$ of the potential is the same as in the hard-hypothesis scenario and that the free energy gain reported in 
equation \ref{soft gain} can 
be a good estimate of the variation of the potential from zero at $r> D+d$ to its maximal absolute value at $r=D$ (i.e. when the beads are in contact). 
Starting from these resonable assumptions we may write
the weighted radial probability distribution as in equation \ref{Boltzman-factor}, by simply substituting the maximal free energy gain of the hard beads scenario
(which is proportional to $3cD/2d$) with that of equation \ref{soft gain} :
\begin{equation}
W_{soft}'(r)dr = W(r) \exp ( c\left(\frac{2(D+d)^{3}-(2^{1/3}D+d)^{3}}{d^{3}}\right)\left(\frac{D+d-r}{d}\right)^2 ) dr
\label{Boltzman-factor-soft}
\end{equation}

 From this expression it is straightforward to obtain the probability distribution of the end-to-end distance, i.e the corrisponding of 
 equations \ref{probability-distribution} 
and \ref{probability-normalized}, and obtain curves analogous to those reported in figure \ref{loop-prob-distribution}.

\subsection{The intron length distribution of higher Eukaryotes}
\label{intron-length}

\begin{figure}[h]
\begin{indented}
\item[]\includegraphics[width=0.8\columnwidth]{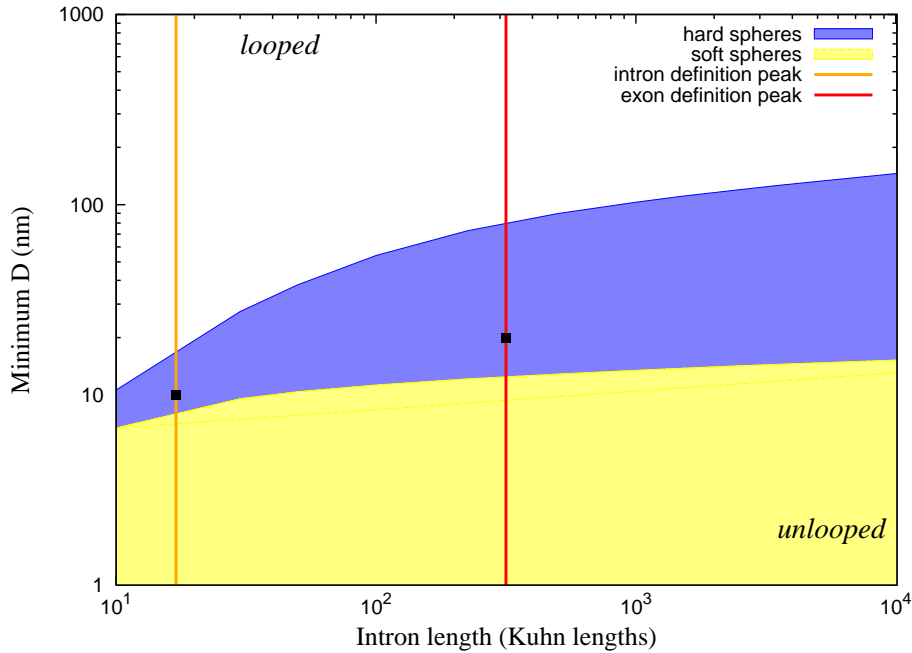}
\end{indented}
\caption{Minimum diameter of beads attached to the ends of an intron needed to obtain a looping probability of 99\%. The two curves (blue and yellow) 
correspond to the
  looping probability  obtained in the hard and soft hypothesis respectively. The two vertical lines (orange and red) correspond to the two peaks in the 
  intron length distribution of H. sapiens (but are quite conserved through different higher Eukaryotes as can be seen in figure \ref{intron-distribution}). 
  Black squares represent the estimated diameter of spliceosome (sub)complexes for the  two corresponding ways of splice-site recognition. While for the 
  intron-definition case estimates for the dimensions of the involved snRNP can be found in literature (\cite{U1,U2}), less information is known for the 
  typical size of the complex contructed across exons in exon definition. In the figure we
  made the (rather conservative) assumption that the diameter of this complex is twice that
  of the subcomplexes involved
in intron-defined splicing.}
\label{cumulative-diameter}
\end{figure}

 If  the depletion attraction plays a role in exon juxtaposition, the typical length of introns with different splicing fate should be in a 
 range suitable to obtain an high looping probability, given the diameter of the  beads attached to their ends.
 In figure \ref{cumulative-diameter} we report the diameter of the beads  needed to have a looping probability of 99\%, in the hard sphere 
 hypothesis (blue line) and soft sphere hypothesis (yellow line). 

To be more precise, the two colored regions represent the D values, obtained by numerical integrations for different intron lengths, for which $P(x<5nm)< 0.99$ 
(see equation \ref{loop-prob}), with the radial probability distributions (described by equation \ref{probability-normalized}), derived starting from 
equation \ref{Boltzman-factor} (hard-sphere) or from equation 
\ref{Boltzman-factor-soft} (soft-sphere).

In figure \ref{cumulative-diameter} we also plot two vertical lines corresponding to the intron lengths
of  the left and right peak of the distribution in figure \ref{intron-distribution} as typical values for the introns devoted to intron definition and exon 
definition respectively. Remarkably enough in both cases the actual dimensions of the splicing complexes (the
black dots along the vertical lines in the figure) lie exactly in between the two
bounds. Moreover looking at the curves it is easy to see  that moving from the first to the second peak, the subcomplexes size must increase
roughly of the amount actually observed in the transition from intron definition to exon definition in
order to keep the same looping probability.

Obviously many other types of specific and elaborate regulation of the splicing dynamic are present in the
   cell, but the ATP-free depletion attraction could explain  the widespread importance of the aspecific intron length variable and the necessity 
of exon definition when the intron length is increased.

Another interesting extension of the model that we propose occurs if the U1 and U2 subcomplexes can form intermolecular bonding. 
In this case there would be an additional force driving the intron looping, besides depletion attraction.
Unfortunately, even if it is likely that such an interaction is present, there is yet no definitive experimental evidence supporting it and, 
what is more important for our purposes,  the nature and form of its potential is still unclear. In particular even the  occurrence of a direct 
interaction is still under debate: while  evidences of such an interaction  were proposed in some early paper  \cite{mattaj,daugeron}, more recent 
works suggest instead that intermediate protein(s) are needed to mediate the interaction.  For instance the need of the protein Prp5 acting as a 
bridge between U1 and U2 was recently discussed in \cite{xu,donmez}. In any case, once the interaction potential will be known, it will be 
rather straigthforward to generalize our model keeping it into account by suitably modifying the Boltzmann factor in equation \ref{probability-distribution}. 
Generally speaking, protein-protein interactions are usually short-range (for example an  hydrogen bond is formed at distances of the order of 0.1-1 nm) and in 
a range of energy compatible with the energy gain due to depletion attraction (see section \ref{presentation}). Thus we may safely predict that an 
additional short-range attraction would only lead to an  overall increasing of the looping probability.  Qualitatively the effect would be a 
left translation of the curves in figure \ref{loop-prob-distribution} and a lowering of the curves in figure \ref{cumulative-diameter}, but 
this would not change the main results of this paper. As a matter of fact only a contribution of the depletion attraction type, introducing a dependence of the looping probability on the diameter of subcomplexes, could explain the switch from intron definition to exon definition.


\subsection{Size constraints on introns and exons}
\label{length-constraints}

\begin{figure}[h]
\begin{indented}
\item[]\includegraphics[width=0.8\columnwidth]{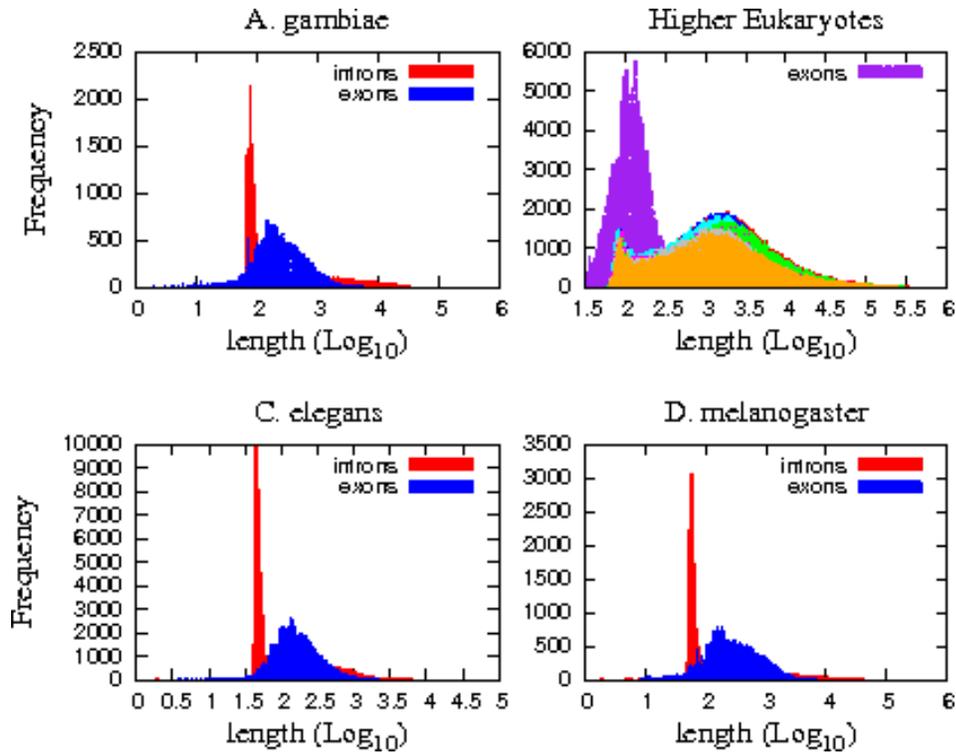}
\end{indented}
\caption{Comparison between intron and exon length distribution in different organisms. The right upper panel represent the superposition of the 
intron length distribution of figure \ref{intron-distribution} with the exon length distribution of the human genome (but this distribution is again 
well conserved through different higher Eukaryotes). In the other three panel the superpositions of intron and exon length distributions for three 
different organisms  (D.melanogaster,A.gambiae,C.elegans) that according to \cite{hertel2}  prevalently use intron definition}
\label{exon-intron}
\end{figure}

Following the idea that the choice of exon or intron definition is related to the looping probability, 
it is expected that organisms which prevalently use intron definition present a strict constraint on their intron 
length but not on their exon length, while the opposite is expected for organisms that prevalently use exon definition.  As reported in many previously 
published studies (\cite{berget,hertel,mcguire,lim} and reference therein) lower Eukaryotes, that prevalently choose intron definition, present a genomic 
architecture typified by small introns with flanking exons of variable length. Higher Eukaryotes have the intron length distribution shown in figure 
\ref{intron-distribution}, with the vast majority of introns devoted to exon definition (see table \ref{tabella}), but a strictly conserved distribution 
of exon length, with a single peack around 100 nt. As shown in the upper right panel of figure \ref{exon-intron}, the position of the typical exon length 
is approximately  the same of the length of introns devoted to intron definition. These values, as discussed above, ensure an high probability of juxtapose 
the two U1 and U2 subcomplexes.\\
 In the case of lower Eukaryotes (three examples in figure \ref{exon-intron}) the intron length distribution presents a single narrow peak in a range compatible 
 with high probability of looping. At the same time no constraint on exons are necessary and indeed the distribution of exons' length is quite broad with a long 
 tail towards large lengths. \\
If the dimensions of merely the U1 and U2 subcomplexes are not enough to ensure an high looping probability across the intron, the exon length is constrained to 
values that give a sufficient looping probability across the exon, allowing the construction of a larger subcomplex that can then lead to the looping of long 
introns, as discussed in the previous section.

\subsection{Cooperative effects}
\label{cooperative}

\begin{figure}[h]
\begin{indented}
\item[]\includegraphics[width=0.8\columnwidth]{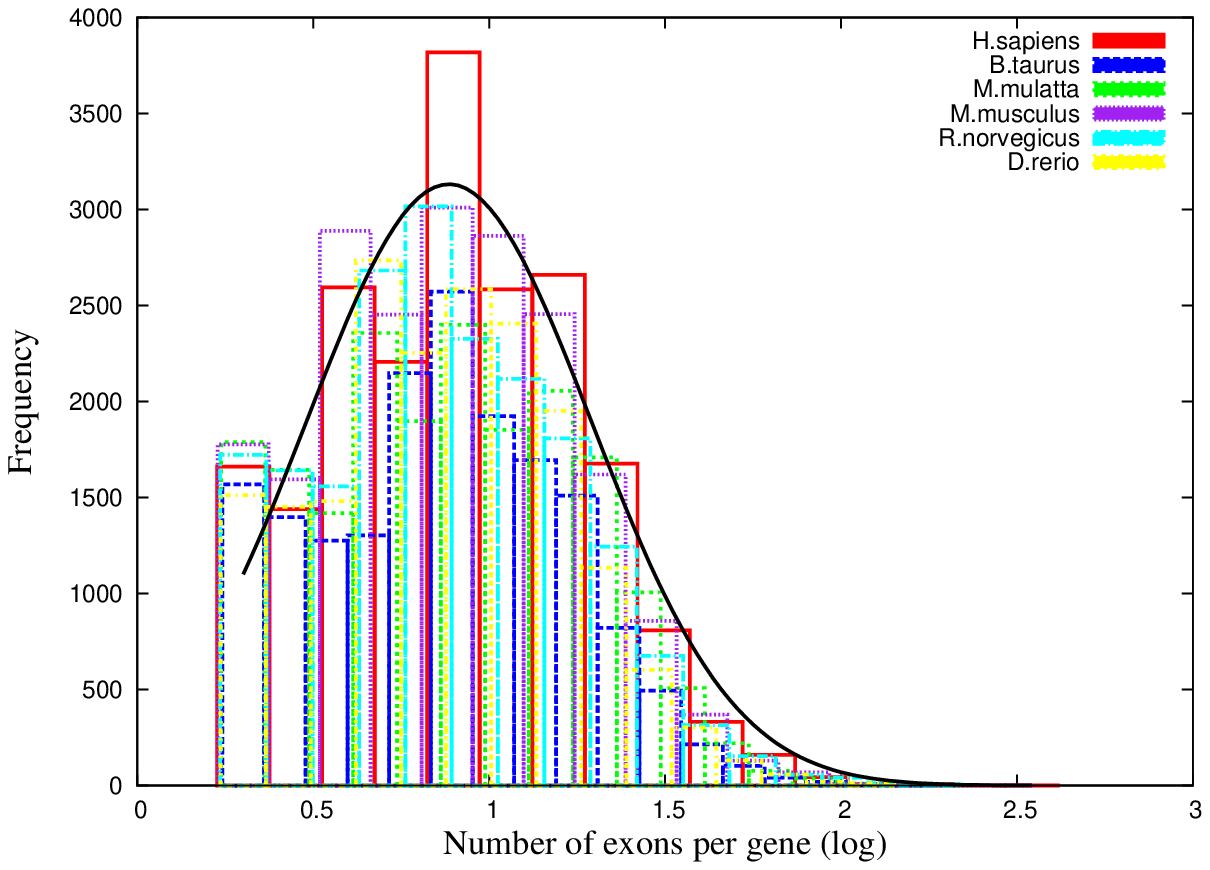}
\end{indented}
\caption{Distribution of the exon number per gene for different higher Eukaryotes (data from \cite{ensembl}).
  Histograms of frequencies are constructed with a logarithmic binning, discarding the intronless genes. The
  continuous line is the result of a tentative gaussian
  fit to the  H.sapiens distribution.}
\label{exon-number}
\end{figure}

So far we completely neglected the cooperative effects that could arise from the presence  of more than two
  beads on the mRNA string. As discussed in \cite{marenduzzo}, the pairing of more than
  two beads moves the energetic balance towards the free energy gain. For example, clustering three beads
  implies three excluded volumes that overlap, but only two loops that have to be closed; four
  beads give a sixfold free energy gain at the cost of closing only three loops, and so forth. However self avoidance
  cannot be neglected in this case, as the increasing number of intron chains progressively makes the looping more energetically expensive. 
As observed in~\cite{marenduzzo} (and reference therein) in three dimensions the entanglement constraints become important when more than eight beads 
cluster together. Above this threshold the free energy gain/loss ratio starts to decrease, setting the optimal number of beads around eight. In the 
framework of exon-defined splicing, each bead corresponds to a
  complex constructed across an exon. Remarkably enough
  the median value of the number of exons per gene is strongly conserved in higher Eukaryotes (which make an
  extensive use of exon-defined splicing) and almost coincides
  with the optimal number of beads in the depletion attraction
  model (see table \ref{tabella} and figure \ref{exon-number}). The same is not true for lower Eukaryotes that  prevalently use intron definition as 
  shown in table \ref{tabella2} for three model organisms.
\begin{table}
     \begin{tabular}{| p{3.5cm}| p{1.5cm}| p{1.5cm}| p{2.5cm}|}
     \hline
     \textbf{Species} &
     \textbf{Median} &
     \textbf{Mean of  the gaussian fit} &
     \textbf{Percentage of  exon-def introns}\\
     \hline
Homo sapiens          &   8  &  7.7 & 84 \\ \hline
Canis familiaris      &   8  &  7.2 & 78 \\ \hline
Pan troglodytes          &   8  &  7.8 & 83 \\ \hline
Danio rerio           &   8  &  6.7 & 66 \\ \hline
Macaca mulatta        &   8  &  6.5 & 79\\ \hline
Mus musculus          &   7  &  6.8 & 84 \\ \hline
Rattus norvegicus     &   7  &  6.6 & 78 \\ \hline
Gallus gallus         &   8  &  6.7 & 83 \\ \hline
Bos taurus            &   8  &  6.8 & 81 \\ \hline

\end{tabular}
\caption{For each species we report: the median (chosen instead of the mean because of the skewness of the distribution) of the overall 
distribution of the number of exons per gene (first column); the mean of the gaussian fit made over the same distribution, discarding the 
intronless genes (second column); the percentage of introns which undergo exon-defined splicing according to \cite{hertel2} (third column).}
\label{tabella}
\end{table}

\begin{table}
     \begin{tabular}{| p{3.5cm}| p{1.5cm}|p{2.5cm}|}
     \hline
     \textbf{Species} &
     \textbf{Median} &
     \textbf{Percentage of  exon-def introns}\\
     \hline
Anophele Gambiae & 3 &  34 \\ \hline
Drosophila melanogaster & 3 &  37 \\ \hline
Caenorhabditis elegans  & 6 &  40 \\ \hline        
\end{tabular}
\caption{For each species we report: the median of the distribution of the number of exons per gene (first column); the percentage of introns 
which undergo exon-defined splicing according to \cite{hertel2} (second column).}
\label{tabella2}
\end{table}

Many more refined and energetically costing mechanisms of splicing are surely present in the cell, and
  many genes present a huge number of exons (up to about 490 in human), but the fact that the typical value
  is mantained in different organism around, or just below, eight, as predicted by the model, seems to suggests
  an evolutionary attempt to mantain the number of beads that maximise the depletion attraction effect in exon juxtaposition.
Our simple modelization does not ensure the joining of exons in the specific order given by the pre-mRNA transcript, allowing the 
possibility of scrumbled exons in the mature mRNA. Despite the fact that  there are several cases reported in literature of this 
scrumbling of exons  \cite{nigro,cocquerelle,shao}, the spliced mRNA usually reproduce the original sequence of exons in the DNA gene, 
eventually with exon skipping or other splicing variations which however do not affect the exons' order. This is probabily due to the 
coupling of splicing to transcription by RNA polymerase \cite{kornblihtt}, which  naturally introduces a polarity in the transcript and 
makes the exons available to the splicing machinery in a sequential manner.

\section{Conclusions and discussion}

We presented a model that highlights the possible role of depletion attraction in the splicing process and we showed that this entropic 
contribution can explain also quantitatively some empirical and bioinformatical observations.
 Spliceosomal introns can perform various functions \cite{roy,federova,zhao,lynch,lynch2} and  the resulting selective forces to mantain 
 or introduce introns during evolution can explain the genome architecture of higher Eukaryotes, characterized by many introns with a typical 
 large size. The necessity to attain a high regulatory capacity within introns can for example explain the average increase of intron size in 
 the mammal branch of the tree of life \cite{pozzoli}. At the same time another splice-site recognition modality has been introduced in higher 
 Eukaryotes: the exon definition. In the perspective of our model the  exon definition pathway was selected by evolution as the simplest way to 
 mantain a balance between the free energy gain due to depletion attraction and the free energy loss caused by looping longer introns. As shown 
 in section \ref{intron-length} the relation between the dimensions of spliceosome subcomplexes and typical intron lengths is in good agreement 
 with our model predictions. With similar arguments we are able to explain the constraints on exons' length: if the  length of introns increases, 
 decreasing their looping probability, the system is compelled to mantain an exons' length suitable for the looping, which is essential to pass to 
 exon definition and obtain diameters of subcomplexes sufficiently large to accomplish the exon juxtaposition.\\
 On the other hand several selective forces can also favour short introns, for example the high fitness of short introns can be due to a reduction 
 of the time and energy cost of transcription and splicing \cite{castillo}, if the conditions favour economy over complexity as in the case of highly 
 expressed genes. Despite the possible selective forces behind - extensively discussed in the case of Drosophila melanogaster in \cite{marais,parsch,halligan}- 
 usually the introns of lower Eukaryotes have been maintained short by evolution. At the same time, there are no evidences of constraints on exon length, a 
 behaviour again perfectly compatible with our model: the complexes on intron boundaries have a dimension which is sufficient to loop the short introns and 
 proceed with the splicing, so no constraint on exons' length is required.
Moreover  evolution led to a proliferation of the number of introns in higher Eukaryotes, leading to the genomes with the  highest density of introns per 
gene \cite{koonin}. This contributes significantly to their proteome complexity \cite{ast,modrek}: a gene with many exons can be spliced in many alternative 
ways to produce different protein products from a single gene. Notwithstanding this, the typical number of exons per gene seems constrained around eight in 
those species that make an extensive use of exon definition. This coincides precisely with the number that allows an optimal exploitation of the depletion 
attraction in exon juxtaposition. This result may suggest a trade off between the advantages of a high number of exons - in terms of complexity - and the 
usage of the uncosting entropic effect of depletion in the splicing process.\\

\ack
This is an author-created, un-copyedited version of an article accepted for publication in Physical Biology.
We thank U. Pozzoli and M.Cereda for very useful discussion and I. Molineris and G. Sales for technical support.
This work was partially supported by the Fund for Investments of Basic Research (FIRB) from the Italian Ministry of the University and Scientific Research, 
No. RBNE03B8KK-006.

\end{document}